\documentclass[utf8]{FrontiersinHarvard} 

\usepackage{url,hyperref,lineno,microtype,subcaption}
\usepackage[onehalfspacing]{setspace}
\usepackage{xcolor}

\def\keyFont{\fontsize{8}{11}\helveticabold }
\def\firstAuthorLast{Yuandong Jia {et~al.}} 
\def\Authors{Yuandong Jia\,$^{1,2}$, Yinbowen Zhang\,$^{2}$, Suwen Wang\,$^{3}$, Guozhi Chai\,$^{2}$, Zemin Zhang\,$^{2}$, Yi Zhang\,$^{4}$, Hongxin Li\,$^{5}$, Shuanglin Huang\,$^{6}$, Hongqing Huo\,$^{2,*}$, Zongfeng Li\,$^{7,*}$ and Yun Kau Lau\,$^{8}$}
\def\Address{$^{1}$School of Physics and Photoelectric Engineering, Key Laboratory of Gravitational Wave Precision Measurement of Zhejiang Province, Taiji Laboratory for Gravitational Wave Universe, Hangzhou Institute for Advanced Study, University of Chinese Academy of Sciences, Hangzhou, China \\

$^{2}$School of Physical Science and Technology, Lanzhou University, Lanzhou, China \\

$^{3}$Hainan Tropical Ocean University, Sanya, China \\
$^{4}$School of Nuclear Science and Technology, Lanzhou University, Lanzhou, China \\
$^{5}$School of Information Science \& Engineering, Lanzhou University, Lanzhou, China \\
$^{6}$Beijing Institute of Mathematical Sciences and Applications, Beijing, China \\
$^{7}$Key Laboratory of Space Utilization, Technology and Engineering Center for Space Utilization, Chinese Academy of Sciences, Beijing, China  \\
$^{8}$Innovation Academy for Microsatellites, Key Lab for Satellite Digitalization Technology, Chinese Academy of Sciences, Shanghai, 201203, China. }
\def\corrAuthor{Hongqing Huo}

\def\corrEmail{huohq@lzu.edu.cn}

\begin{document}
\onecolumn
\firstpage{1}

\title[UV-LEDs and charge management]{Long-wavelength UV-LEDs and charge management in the detection of gravitational waves in space} 

\author[\firstAuthorLast ]{\Authors} 
\address{} 
\correspondance{} 

\extraAuth{Zongfeng Li \\ lzfeng@csu.ac.cn}

\maketitle

\begin{abstract}

\section{}
For the charge management system in gravitational wave detection missions, a continuous discharge strategy is considered by continuously illuminating a test mass (TM) with weak light in such a way to strike a balance between the charging and discharging rates and at the same time avoids the requirement for frequent activation of charge measurements. Built on experiments by one of us \cite{TATP} based on a simple parallel plate model for inertial sensor, in the present work a  more sophisticated inertial sensor model  that mimics the surface properties and work function of a cubical TM of an inertial sensor in space (like that of the LISA Pathfinder) is employed to study bipolar charge management system that utilizes UV-LEDs with peak wavelengths of 269 nm, 275 nm, 280 nm, and 295 nm that are longer than the standard 255 nm commonly employed for direct TM illumination.  Experimental results indicate that the 275 nm UV-LED achieves optimal performance, maintaining the TM potential closer to zero and at the same time accommodates both rapid discharge and continuous discharge strategies. The present work provides useful input in the future study of system design and optimization for the charge management system.

\tiny
 \keyFont{ \section{Keywords:} UV-LED, charge management, inertial sensors, passive charge control, discharge strategy} 
\end{abstract}

\section{Introduction}

In gravitational wave detection in space, spacecraft is exposed to the heliospheric environment and subjected to continuous bombardment by high energy particles \cite{ECST, CFFT, DCTC}. When the energies exceed a certain threshold, particles will penetrate the spacecraft and deposit on a TM, generating spurious acceleration noise and disturbing gravitational wave detection \cite{CDSL, CIAN}. Beginning from the Gravity Probe B(GP-B) mission and later on the LISA Pathfinder(LPF), photoelectric effect is employed as a non-contact way to neutralise charges where the mercury lamp is used as the ultraviolet(UV) light sourse \cite{CMCT, LLPC, TTUD, CIFN}. Instead of mercury lamp, UV-LED is considered as a better alternative for a number of technical reasons in the upcoming LISA and the prospective Chinese missions, such as TaiJi and TianQin \cite{ULLT, TTPS, TASG}. UV-LED based charge management systems are now widely studied, and Stanford University has successfully demonstrated charge management system with UV-LEDs in space \cite{GTFD, CTDU, ACCM, TATP}.  

During a solar cycle, charge control strategies vary with particle flux intensity. Among them, continuous discharge is used to balance the charging-discharging rates, typically suited for the ascending and descending phase of a solar cycle in which particle flux is predominantly due to galactic cosmic rays and the solar activity is low. A continuous discharge method employing dual UV light sources was proposed to illuminate both the TM and the electrode housing, effectively maintaining the TM potential near zero \cite{CCMS, TATP, PCCI}. Previously, Wang and the Stanford group studied continuous charge management methods by employing UV light with a peak wavelength longer than the 255 nm commonly used, and they used a simplified parallel plate model derived from an inertial sensor \cite{TATP}. 

The aim of the present work is to conduct this study by employing a more realistic model of an inertial sensor instead of a simple parallel plate model. A cubic TM is fixed within an electrode housing via insulating pillars and the surfaces are gold coated to mimic those of the LPF in terms of the work function and surface properties of the TM as well as the electrode housing. Continuous discharge requires sustained operation during solar minimum periods. However, constrained by the experimental conditions of this study, we use 4-hour continuous discharge results to  demonstrate in a preliminary way its performance, corresponding to the 0.1 mHz frequency band essential for space-based gravitational wave detection. A preliminary continuous charge management is then studied using multiple wavelengths of UV-LEDs. Experimental results indicate that, due to internal reflections of UV light within the electrical housing, the photoelectron emission is different from that of a parallel plate model. Further, our results show that 275 nm outperforms the standard 255 nm wavelength for both fast release of charged particles as well as for continuous charge management, it is capable of  achieving smaller equilibrium potential with long time stability. 

In the present work,  we investigate the UV-LED light sources for charge management. Although these light sources have been extensively studied by a number of research groups with focus on the 255 nm wavelength. Our contribution lies in exploring longer wavelength light sources beyond conventional 255 nm, and we argue that during periods of low solar activity, the 275 nm light source seems to be a better option. This result is similar to our previous experiments based on UV micro-LED light sources(\cite{TMCM}), where we used four UV micro-LEDs of different wavelengths. Light sources with wavelengths longer than the 255 nm commonly used achieved a equilibrium potential closer to 0 mV, with 274 nm reaching -10 mV. However, in the paper(\cite{TMCM}), we focus more on the feasibility of applying this new light source to charge management in the detection of gravitational wave in space. We conducted a characterization analysis of the new light source and verified its potential application in the space environment.

The present work is structured as follows. In section 2, charge management model based on a UV-LED with long wavelength is introduced and analyzed. In section 3, based on the cubic charge management system, photoelectric effect for a UV-LED with peak wavelength of 275 nm is demonstrated. Continuous charge control using UV-LEDs with different peak wavelengths will be studied in section 4, which will enable us to explore a method using light sources different from the standard 255 nm.  The experimental results show that UV-LEDs with longer wavelengths achieve excellent long-term stability, maintaining equilibrium potentials within approximately 50 mV over a four-hour period, and achievng a charge noise level of $\rm 10^{-13}\ C/ \sqrt{Hz}$  near 0.1 mHz. Moreover, 275 nm UV-LED achieves not only smaller equilibrium potential (-12 mV) but also faster discharge rate (6 s), making it a promising candidate as a better light source for charge management system.

\section{Principle of charge management}

In this section, we will present a  charge management model for long-wavelength UV-LED illumination, on the basis of which we try to understand our experimental results and perform data analysis. 

\begin{figure}[h!]
\begin{center}
\includegraphics[width=10cm]{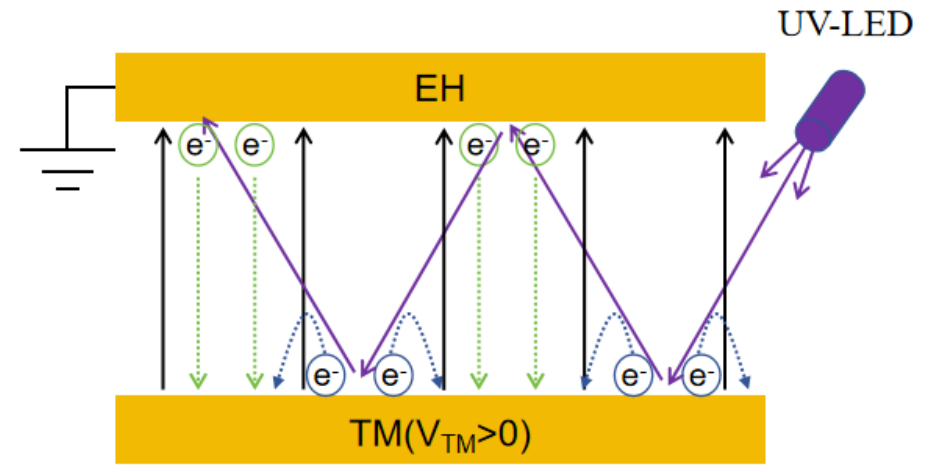}
\end{center}
\caption{ A schematic of photoelectron migration when the potential of the TM is positive and greater than the initial kinetic energy of the photoelectrons.}\label{prin1}
\end{figure}

Fig. \ref{prin1} presents a schematic representation of the charge management mechanism employing long-wavelength (i.e., wavelength longer than 255 nm) UV-LED for direct illumination. The work function of Au exposed to air is approximately 4.2 eV \cite{ISSP, PFGT}. As a result, incident light with a photon energy greater than 4.2 eV (corresponding to a wavelength of less than ~295 nm) is sufficient to induce photoemission for charge management. In this configuration, UV light illuminates the TM and reflects between the TM and the electrode housing (EH). Photoelectrons are subsequently emitted from both gold-coated surfaces. 

\begin{figure}[h!]
\begin{center}
\includegraphics[width=10cm]{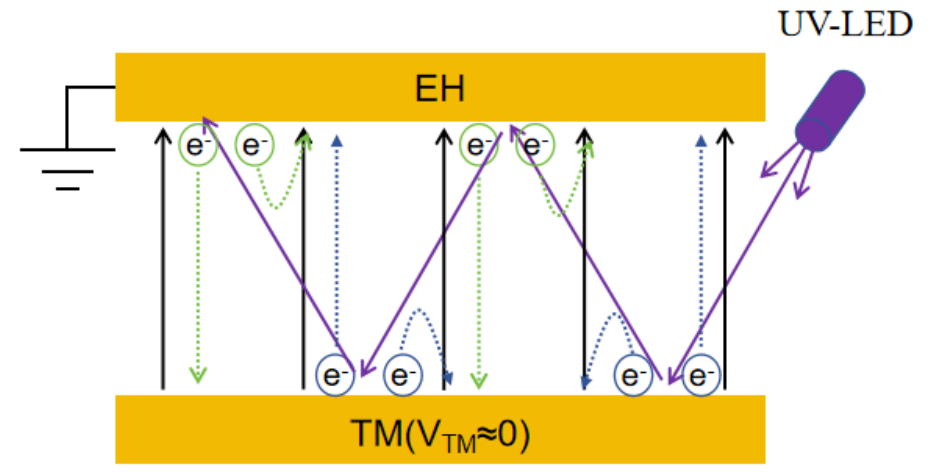}
\end{center}
\caption{ A schematic of photoelectron migration when the potential of the TM approaches zero and becomes comparable to the initial kinetic energy of the photoelectrons}\label{prin2}
\end{figure}

Consider the case in which the potential of the TM is positive and greater than the initial kinetic energy of the photoelectrons, where the electrode housing (including its mounted electrodes) is grounded. the presence of the local electric field causes photoelectrons to migrate to the TM, thereby reducing its potential, as shown in Fig. \ref{prin1}. When the potential approaches zero and becomes comparable to the initial kinetic energy of the photoelectrons, the photoelectron flow between the TM and the electrode housing reaches an equilibrium state. At this point, the TM attains a stable potential, as illustrated in Fig. \ref{prin2}. A similar process occurs when the initial TM charge is negative. As the wavelength of the incident light increases, the photon energy decreases, which correspondingly reduces the kinetic energy of the generated photoelectrons. Thereby, a smaller equilibrium potential of the TM can be achieved.

The application of a bias voltage ($V_{\rm B}$) to the electrode will influence the migration of photoelectrons between the TM and the electrode housing. The potential variation of the TM under UV illumination follows an exponential function, which has been verified through both theoretical modeling and experimental validation \cite{CCMS, MEUD}: 

\begin{equation}
\label{eq:1}
V_{\rm TM} \left ( t \right ) =A\mathrm{e} ^{-\frac{t}{\tau } }+V_{\rm eq}(V_{\rm B})  
\end{equation}

\noindent
where $\tau$ is the time constant that reflects the potential variation rate of the TM, and $V_{eq}(V_{\rm B})$ is the equilibrium potential to which the TM will settle under a given bias voltage after prolonged illumination. 

\section{Ground experiments of charge management.}
\subsection{Experimental setup}
In this work, instead of the parallel plate model, we conducted experiments using a cubic TM enclosed in an electrical housing, as in the case of the LISA Pathfinder mission. This enables us to evaluate the  performance of UV-LEDs in a setting that closely resembles that on orbit.

Fig. \ref{sch&graph} illustrates the experiment setup for validating long-wavelength UV-LED based charge management systems. The vacuum chamber maintained an operational pressure on the order of $10^{-4}$ Pa throughout testing. The electrical housing model was designed to mimic the inertial sensor of  LISA Pathfinder, featuring a measured TM-to-surroundings capacitance of 28.8 pF. 
Both the surface of the TM and the inner surface of the electrode housing were gold-coated to a thickness of 500 nm \cite{ULCC}. The TM was fixed inside the electrode housing using insulated Ultem-1000 holding tubes, which provided a TM-to-EH resistance on the order of $10^{14}\ \Omega$. UV-LEDs were mounted at the upper vertices of the housing, providing direct illumination on the TM. A contact probe mounted at the centre of the upper surface of the TM was used for real-time potential measurements with an accuracy of 0.1 mV, while the bias electrode was either grounded or connected to an external voltage source to provide a bias potential.  This configuration deviates from the previous simple two parallel plates model and allows for a comprehensive analysis of the charge management process under conditions that closely mimic those encountered in actual space missions.

\setcounter{figure}{3}
\setcounter{subfigure}{0}
\begin{subfigure}
\setcounter{figure}{3}
\setcounter{subfigure}{0}
    \begin{minipage}[b]{0.5\textwidth}
        \includegraphics[width=\linewidth]{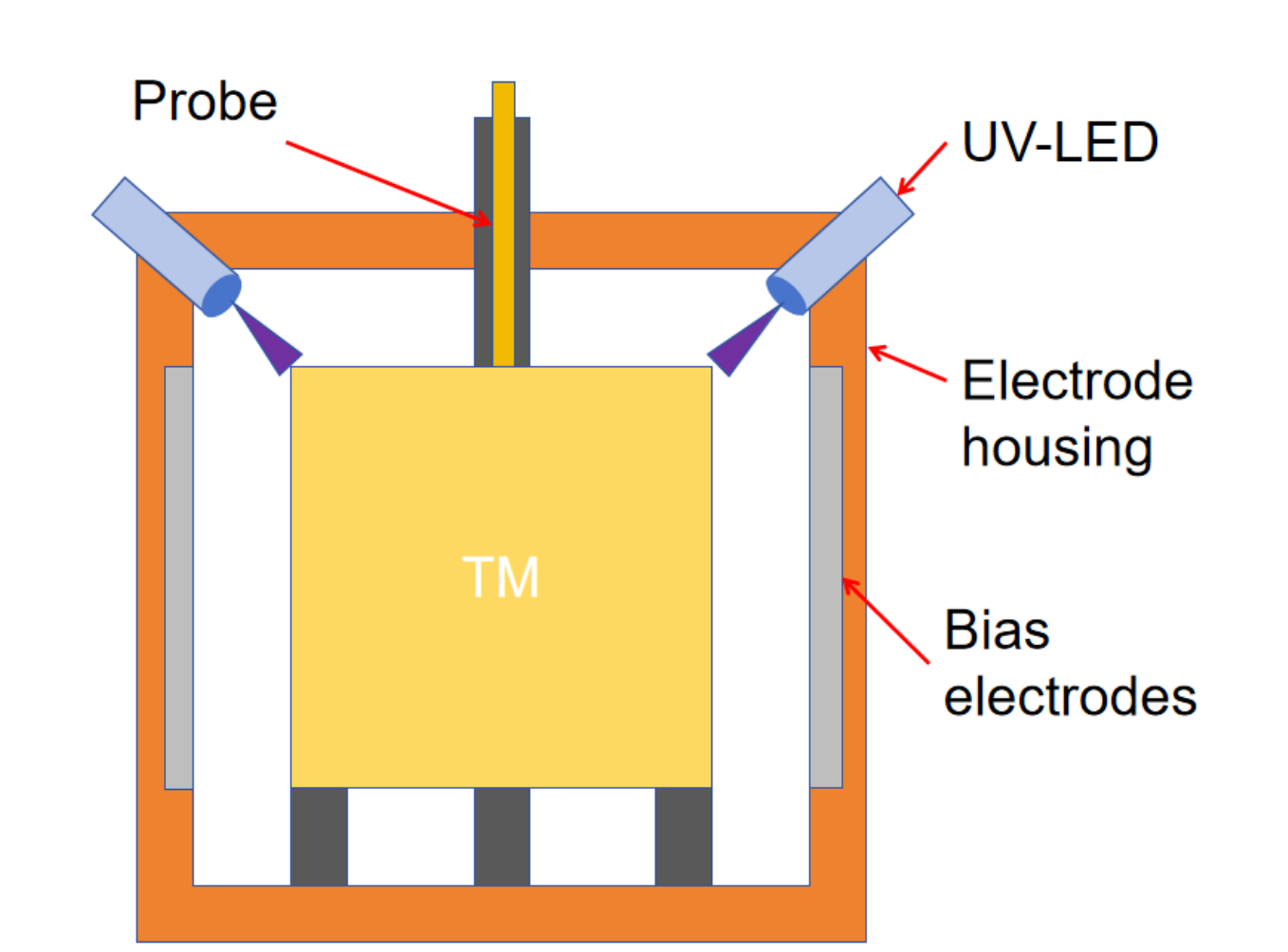}
        \caption{Schematic.}
        \label{sch}
    \end{minipage}  
\setcounter{figure}{3}
\setcounter{subfigure}{1}
    \begin{minipage}[b]{0.5\textwidth}
        \includegraphics[width=\linewidth]{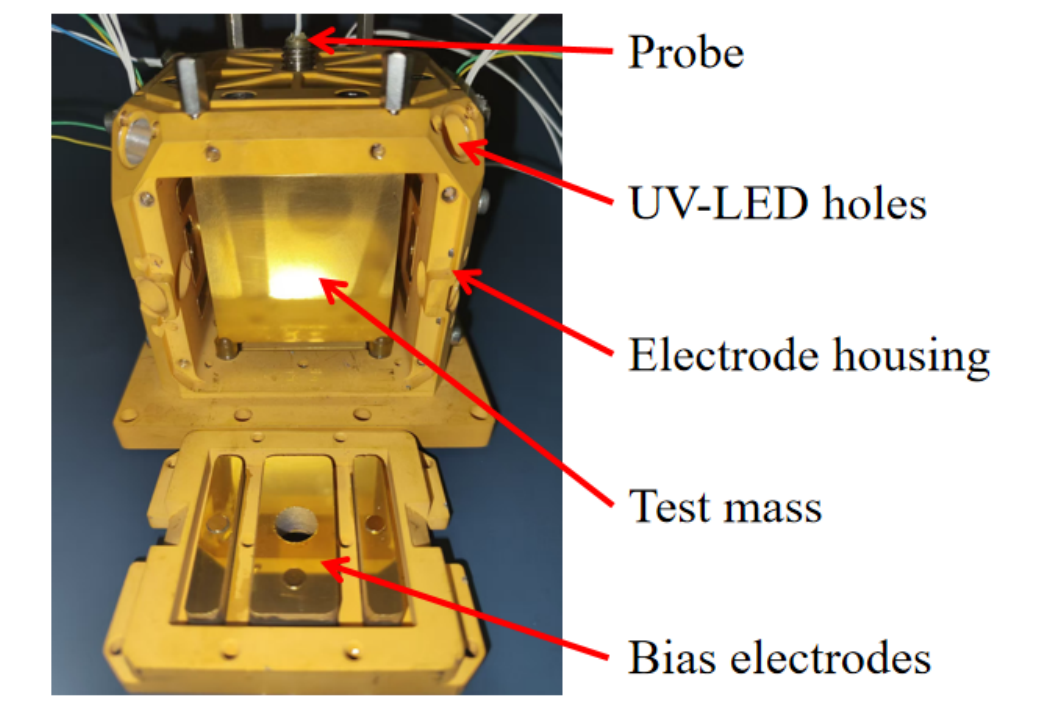}
        \caption{Photograph.}
        \label{graph}
    \end{minipage}
\setcounter{figure}{3}
\setcounter{subfigure}{-1}
    \caption{Experiment setup of the charge management system. \textbf{(A)} Schematic. \textbf{(B)} Photograph.}
    \label{sch&graph}
\end{subfigure}

Fig. \ref{LED_chara} presents the emission spectra and optical power versus current (P-I) of the five UV-LEDs used in this study. During the P-I curve measurements, the power meter was positioned at a distance of 3 cm from the UV-LEDs. Table \ref{wavelength} summarizes the peak wavelengths and full width at half maximum (FWHM) values for all five UV-LEDs.

\setcounter{figure}{4}
\setcounter{subfigure}{0}
\begin{subfigure}
\setcounter{figure}{4}
\setcounter{subfigure}{0}
    \begin{minipage}[b]{0.5\textwidth}
        \includegraphics[width=\linewidth]{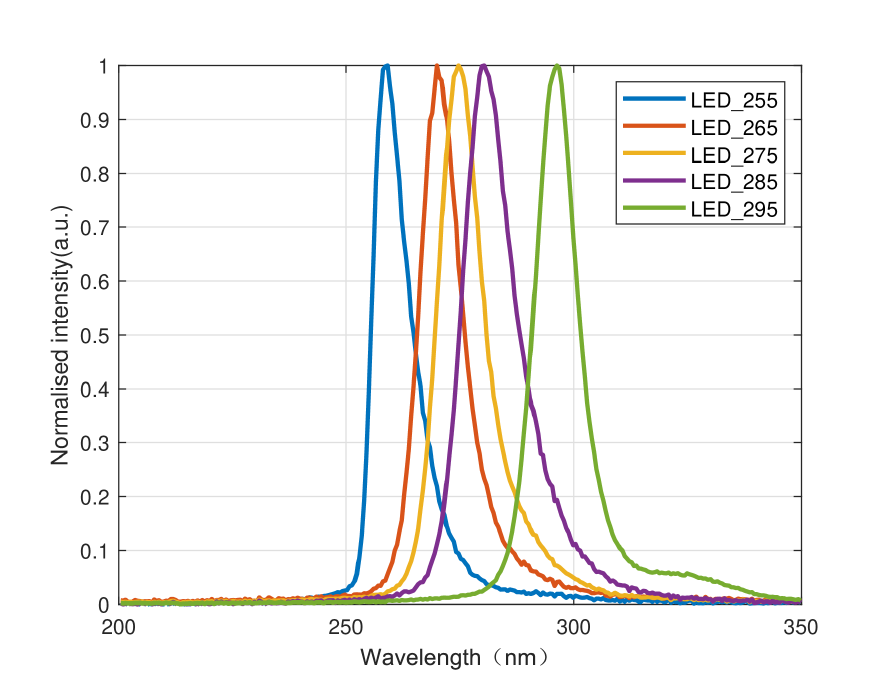}
        \caption{Emission spectra of the UV-LEDs.}
        \label{spectrum}
    \end{minipage}  
\setcounter{figure}{4}
\setcounter{subfigure}{1}
    \begin{minipage}[b]{0.5\textwidth}
        \includegraphics[width=\linewidth]{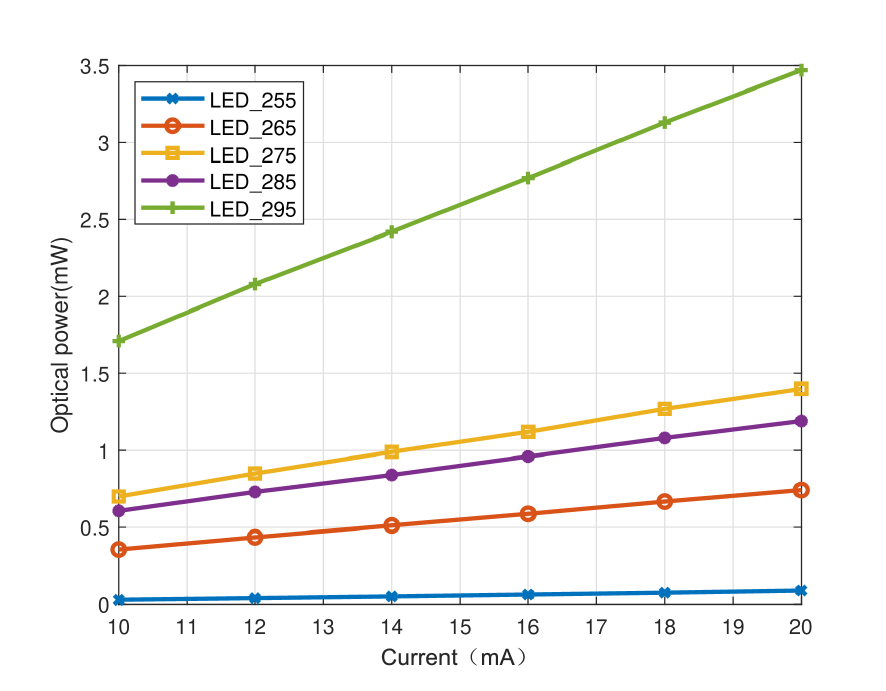}
        \caption{P-I curve of the UV-LEDs.}
        \label{IP}
    \end{minipage}

\setcounter{figure}{4}
\setcounter{subfigure}{-1}
    \caption{Characteristic performance of the five UV-LEDs in the experiment. \textbf{(A)} Emission spectra. \textbf{(B)} P-I curve.}
    \label{LED_chara}
\end{subfigure}

\begin{table}
\caption{Characteristics of the UV-LEDs used in the experiment}
\label{wavelength}
\begin{tabular*}{\textwidth}{@{\extracolsep{\fill}}ccccc}
\hline
\textbf{LED}    &\textbf{Peak wavelength(nm)}   &\textbf{FWHM(nm)}  &\textbf{Manufacturer}  &\textbf{Part number}\\
\hline
LED\_255	&$259.1 \pm 0.5$	&$9 \pm 1$  &NewOpto    &UVLED255-TO39HS\\
LED\_265	&$269.4 \pm 0.5$	&$10 \pm 1$ &NewOpto    &UVLED265-TO39HS\\
LED\_275	&$274.6 \pm 0.5$	&$11 \pm 1$ &NewOpto    &UVLED275-TO39HS\\
LED\_285	&$280.3 \pm 0.5$	&$13 \pm 1$ &NewOpto    &UVLED285-TO39HS\\
LED\_295	&$296.3 \pm 0.5$	&$11 \pm 1$ &NewOpto    &UVLED295-TO39HS\\
\hline
\end{tabular*}
\end{table}

\subsection{Photoelectrons emission.}

To study the charge control capability of the system, two experimental runs were conducted using the LED\_275. In these experiments, an increase in the potential of the TM was defined as charging, while a decrease was defined as discharging. Fig. \ref{bias} illustrates the correlation between the bias electrode potential ($V_{\rm B}$) and the TM potential ($V_{\rm TM}$), while the UV-LED drive current was maintained at 10 mA during the measurements.

\setcounter{figure}{5}
\setcounter{subfigure}{0}
\begin{subfigure}
\setcounter{figure}{5}
\setcounter{subfigure}{0}
    \begin{minipage}[b]{0.5\textwidth}
        \includegraphics[width=\linewidth]{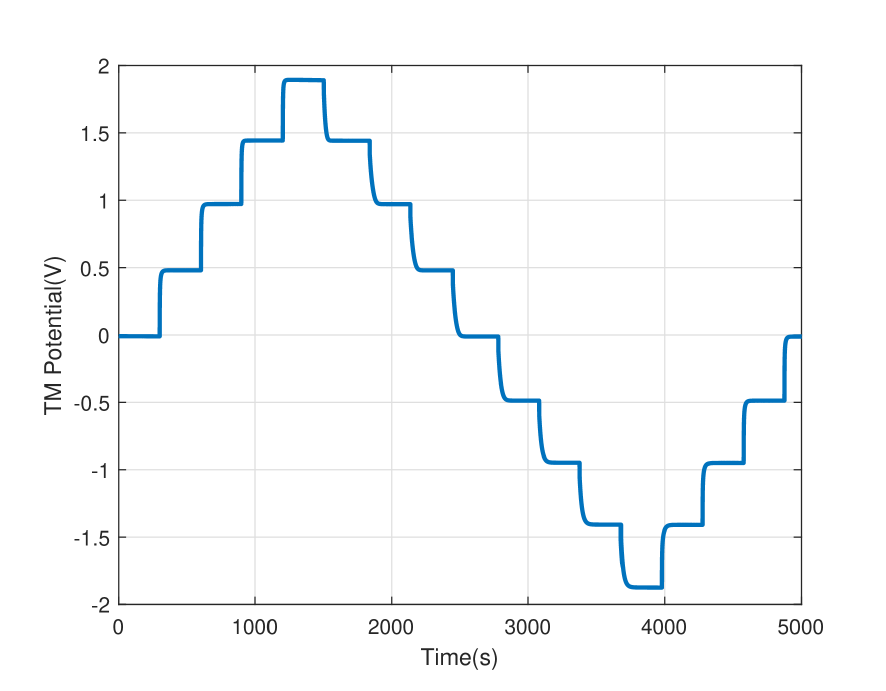}
        \caption{$V_{\rm TM}$ as a function of time with $V_{\rm B}$.}
        \label{bias1}
    \end{minipage}  
\setcounter{figure}{5}
\setcounter{subfigure}{1}
    \begin{minipage}[b]{0.5\textwidth}
        \includegraphics[width=\linewidth]{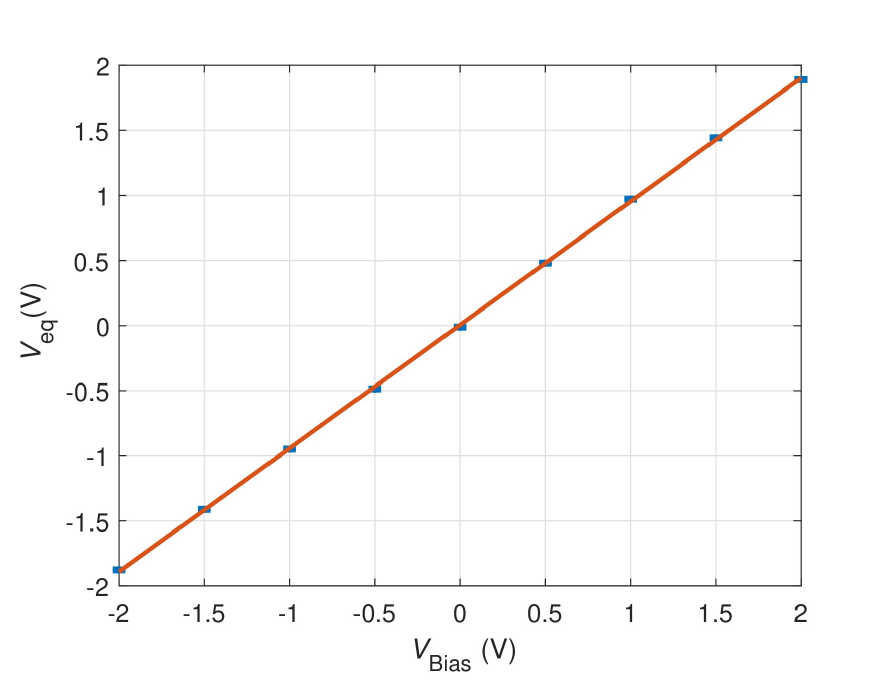}
        \caption{$V_{\rm eq}$ as a function of $V_{\rm B}$.}
        \label{bias2}
    \end{minipage}

\setcounter{figure}{5}
\setcounter{subfigure}{-1}
    \caption{Demonstration of the variation of $V_{\rm TM}$ affected by $V_{\rm B}$ under the illumination of LED\_275. \textbf{(A)} $V_{\rm TM}$ as a function of time with $V_{\rm B}$ varied from -2 V to 2 V in steps of 0.5 V. \textbf{(B)} $V_{\rm eq}$ as a function of $V_{\rm B}$.}
    \label{bias}
\end{subfigure}

A continuous test was conducted by sweeping $V_{\rm B}$ from 0 V to +2 V, then to -2 V, and finally returning to 0 V in steps of 0.5 V. Each voltage setting was maintained for 300 seconds while monitoring $V_{\rm TM}$ until reaching equilibrium (Fig. \ref{bias}a). Based on (\ref{eq:1}), the equilibrium potentials of the TM $V_{\rm eq}$ corresponding to nine discrete $V_{\rm B}$ configurations are plotted in Fig. {\ref{bias}b}, where the solid line represents the fit to the data:

\begin{equation} 
\label{eq:fit1}
{V_{\rm eq}\left (\rm V\right )} = \left (0.948 \pm 0.004\right ) \times {V_{\rm B}\left (\rm V\right ) + \left (0.005 \pm 0.005\right ) \left (\rm V\right )}
\end{equation}

The $R^2$ of the fitting is 0.99989, the slope is 0.948 with a 95\% confidence interval (CI) of [0.939, 0.957], and the intercept is 0.005 V with a 95\% CI of [-0.006, 0.017] V.

The charge management system performance was also influenced by the UV-LED drive current. In this experiment, the LED\_275 drive current was set to 10 mA, 15 mA and 20 mA respectively while the bias electrode was either grounded or set to $V_{\rm B} = $ 1 V. The corresponding TM potential variations are presented in Figure \ref{currentp}.
The experimental results reveal an inverse relationship between the UV-LED drive current and the stabilization time of the TM potential, while the equilibrium potential is independent of it. Table \ref{current} summarizes the maxima variation of TM potential at the start of discharge $(\mathrm{d} V/\mathrm{d}t)^{-}$ and charging $(\mathrm{d} V/\mathrm{d}t)^{+}$ with about 5\% uncertainty for the three sets of drive current. It should be noted that the discharge rate differs from the charging rate by a factor of about 30. This discrepancy arises from two main aspects: first, the net photoelectron flux between the TM and the electrode housing depends on the potential polarity of the TM; second, it is influenced by the difference between the initial potential and the equilibrium potential of the TM.
\setcounter{figure}{6}
\setcounter{subfigure}{0}
\begin{subfigure}
\setcounter{figure}{6}
\setcounter{subfigure}{0}
    \begin{minipage}[b]{0.5\textwidth}
        \includegraphics[width=\linewidth]{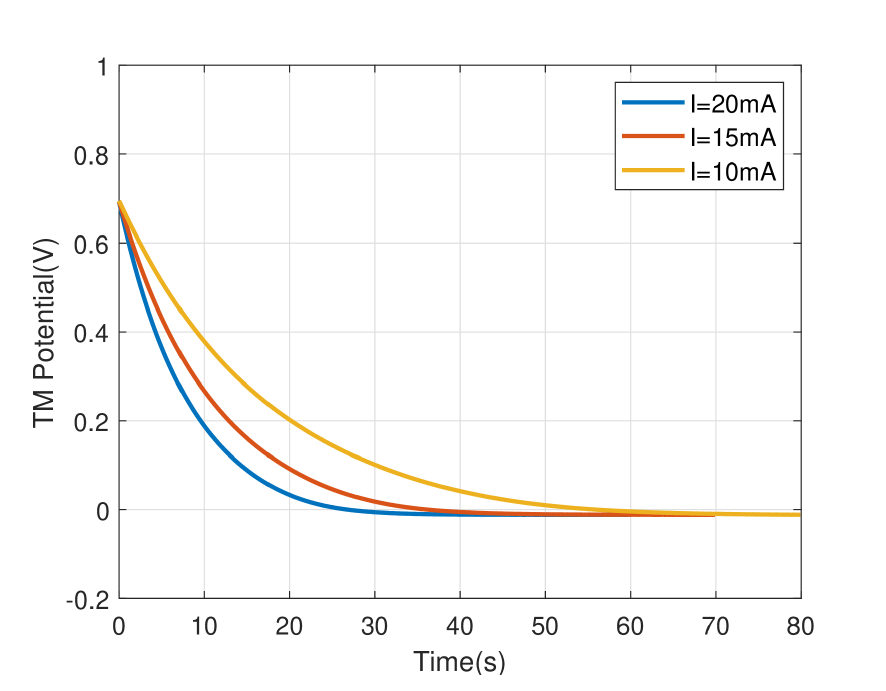}
        \caption{The bias electrode was grounded.}
        \label{B0}
    \end{minipage}  
\setcounter{figure}{6}
\setcounter{subfigure}{1}
    \begin{minipage}[b]{0.5\textwidth}
        \includegraphics[width=\linewidth]{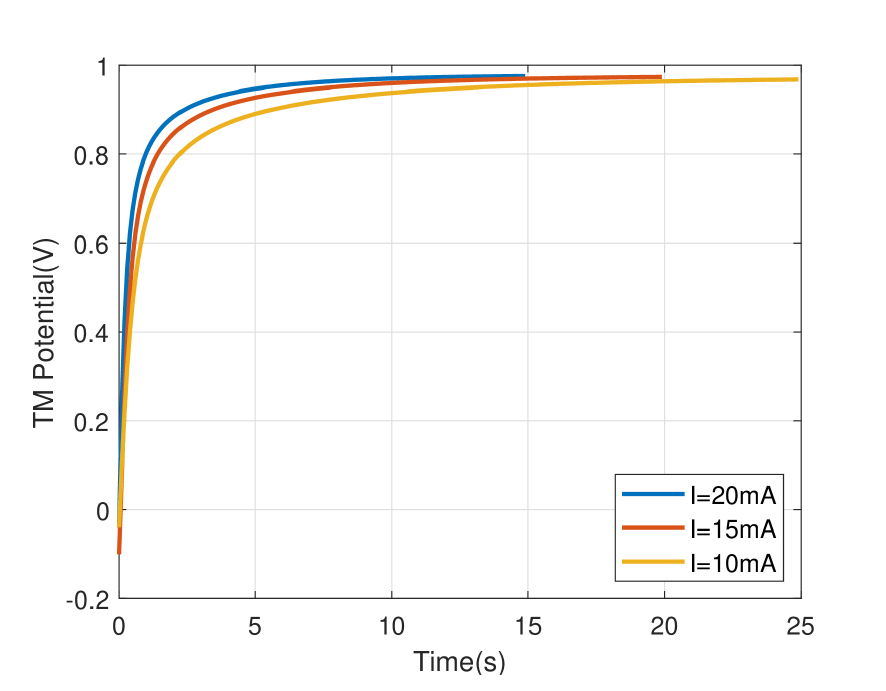}
        \caption{$V_{\rm eq}$ as a function of $V_{\rm B}$.}
        \label{B1}
    \end{minipage}

\setcounter{figure}{6}
\setcounter{subfigure}{-1}
    \caption{TM potential variation for LED\_275 with drive currents of 10 mA, 15 mA, and 20 mA. \textbf{(A)} The bias electrode was grounded. \textbf{(B)} The potential of bias electrode was set to 1 V.}
    \label{currentp}
\end{subfigure}

\begin{table}
\caption{Maxima of potential variations under different currents for the charging and discharging from Fig. \ref{currentp}.\label{current}}
\begin{tabular*}{\textwidth}{@{\extracolsep{\fill}}cccccc}
\hline
\textbf{I(mA)}	& \textbf{$(\rm{d} V/\rm{d}t)^{-}(\rm{mV/s})$}	& \textbf{$I(\rm pA)$} &
\textbf{$(\rm{d} V/\rm{d}t)^{+}(\rm{mV/s})$} &
\textbf{$I(\rm pA)$}\\
\hline
10		& 84.5  & 2.43		& 2314	& 66.6\\
15		& 61.6	& 1.78		& 1793	& 51.6\\
20		& 39.5	& 1.13		& 1290	& 37.2\\
\hline
\end{tabular*}

\end{table}

\section{Continuous charge control.}

In this experiment, long-wavelength UV-LEDs were used to evaluate the charge management performance. The performance was quantified by three metrics: the equilibrium potential of the TM, the stability of this potential, and the discharge rate.

\begin{figure}[h!]
\begin{center}
\includegraphics[width=10cm]{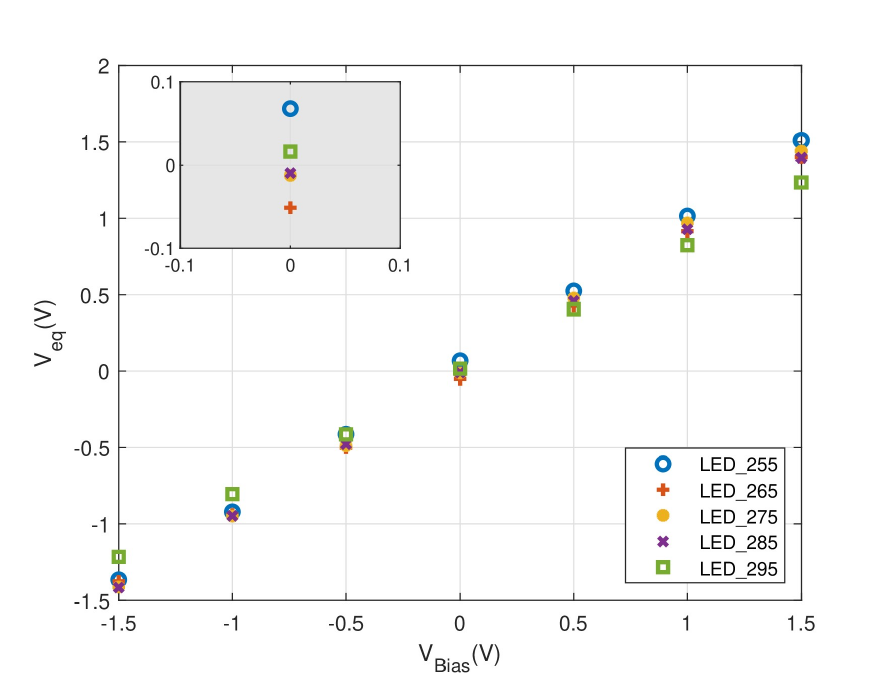}
\end{center}
\caption{ TM equilibrium potential as a function of the bias potential for different wavelength UV-LEDs}\label{multi1}
\end{figure}

Fig. \ref{multi1} illustrates the equilibrium potential $V_{\rm eq}$ as a function of the bias electrode potential $V_{\rm B}$ under illumination by UV-LEDs of five different wavelengths. The zero-bias regime ($V_{\rm B}$ = 0 V) represents the primary focus of this study, with its detailed characteristics highlighted in the enlarged inset in the upper-left corner.

The equilibrium potential of the TM under zero-bias conditions ($V_{\rm B}$ =  0 V) was used to quantify the charge management performance. Table \ref{multi1t} summarizes these measurements (corresponding to the shaded insert in figure \ref{multi1}), demonstrating that direct TM illumination enables continuous charge control. The results show that all five UV-LEDs maintain the TM charge within $\pm$100 mV, the LISA requirement (eqivalent to about $2 \times 10^7$ elementary charges \cite{IBSE}. It should be noted that LED\_255 and LED\_265 achieve comparable equilibrium potentials around 50 mV, while the longer LED\_275, LED\_285, and LED\_295) are moving the TM significantly closer to zero potential, indicating superior performance in precision charge management. It should be noted that the polarity of the equilibrium potential varies under different UV-LEDs. This is because the equilibrium potential is governed by the net photoelectron flux between the TM and the electrode housing, which in turn is influenced by factors such as the incident light wavelength, the geometry of the electrode housing, and the physical properties of the gold-coated surfaces. Therefore, further investigation through simulation is required to fully understand the behaviour of the equilibrium potential.

\begin{table}
\caption{TM equilibrium potential under illumination by UV-LEDs of different wavelengths.\label{multi1t}}

\begin{tabular*}{\textwidth}{@{\extracolsep{\fill}}cc}
\hline
\textbf{LED}	& \textbf{Equilibrium potential(mV)}	\\
\hline
LED\_255		& $68.4 \pm 0.6$	\\
LED\_265		& $-51.5 \pm 0.2$	\\
LED\_275		& $-12.0 \pm 0.2$	\\
LED\_285		& $-9.9 \pm 0.8$	\\
LED\_295		& $16.1 \pm 1.1$	\\
\hline
\end{tabular*}

\end{table}

To evaluate the temporal stability of the long-wavelength UV-LED-based charge management system, four-hour continuous tests were conducted using LED\_265, LED\_275, LED\_285 and LED\_295. During these tests, the drive current was set to 10 mA, and the bias electrode was grounded. Fig. \ref{long}(a) illustrates $V_{\rm TM}$ over time for each wavelength, and Fig. \ref{long}(b) shows the equivalent TM charge noise under the control of the LED\_275.

\setcounter{figure}{8}
\setcounter{subfigure}{0}
\begin{subfigure}

\setcounter{figure}{8}
\setcounter{subfigure}{0}
    \begin{minipage}[b]{0.5\textwidth}
        \includegraphics[width=\linewidth]{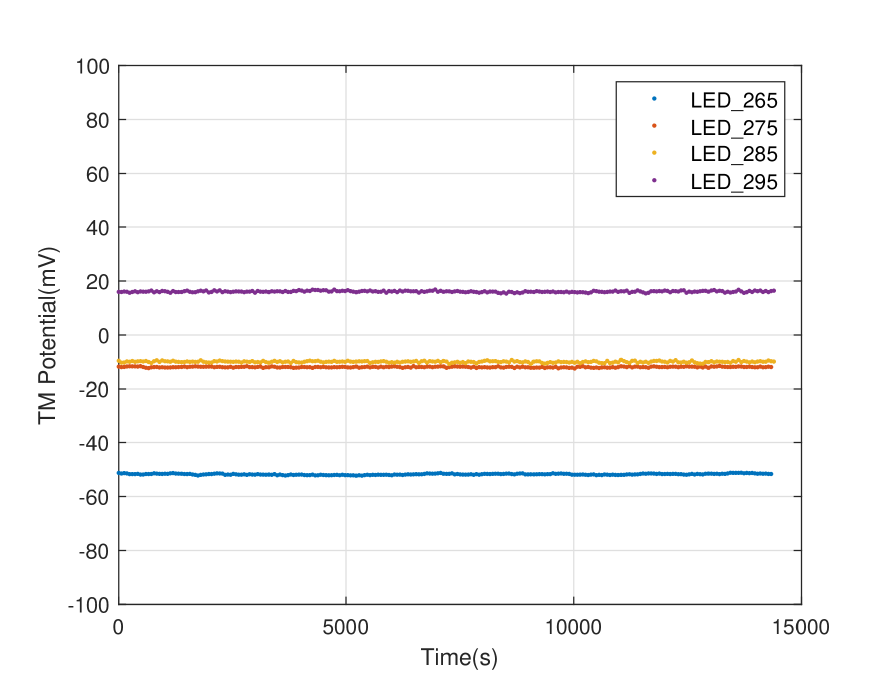}
        \caption{TM potential drift about 4 hours.}
        \label{long1}
    \end{minipage}  
\setcounter{figure}{8}
\setcounter{subfigure}{1}
    \begin{minipage}[b]{0.5\textwidth}
        \includegraphics[width=\linewidth]{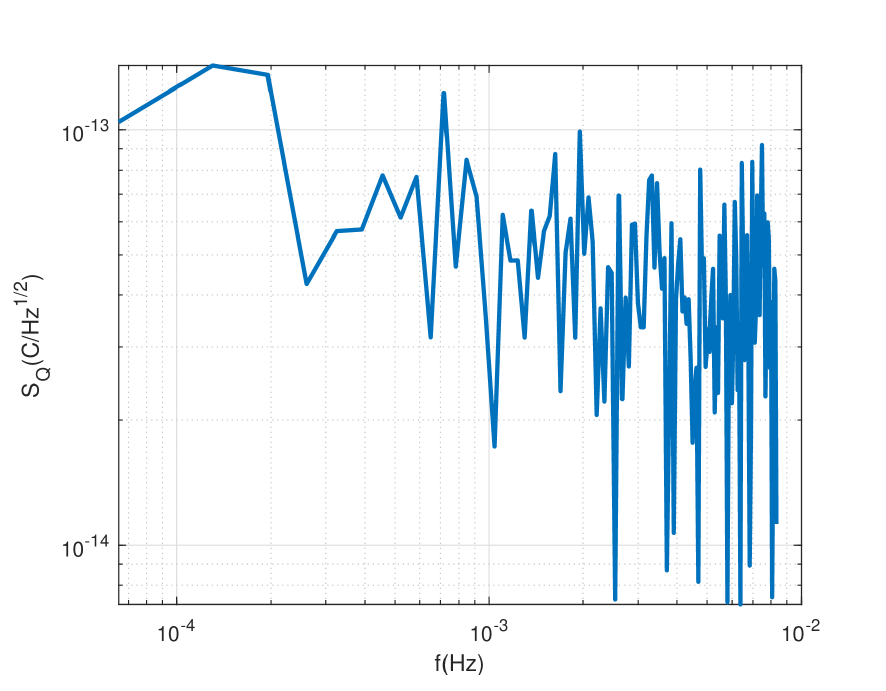}
        \caption{$V_{\rm eq}$ as a function of $V_{\rm B}$.}
        \label{long2}
    \end{minipage}

\setcounter{figure}{8}
\setcounter{subfigure}{-1}
    \caption{Experiment results of long-wavelength UV-LED-based charge management system. \textbf{(A)} TM potential drift about 4 hours. \textbf{(B)} TM charge noise under the control of the LED\_275.}
    \label{long}
\end{subfigure}

The experimental results demonstrate stable maintenance of the TM potential within $\pm$100 mV of the target value over the 4-hour duration, with all four wavelengths exhibiting peak-to-peak fluctuations below 2 mV. This confirms the viability of long-wavelength UV-LEDs for passive charge management in gravitational reference sensor applications. Under the control of the LED\_275, the system achieved a charge noise level of $\rm 10^{-13}\ C/ \sqrt{Hz}$  near 0.1 mHz, corresponding to an equivalent acceleration noise of $10^{-15} \ \rm{m \cdot s^{-2} / \sqrt{Hz}}$ as derived from the literature \cite{CIAN}. Subsequently, we will conduct longer-duration continuous discharge experiments, including tests at lower frequency bands with both charge management on and charge management off, thereby validating its capability for sustained operation throughout the entire mission duration. And the 4-hour test duration is relatively brief compared to a full space mission cycle, and additional risks may exist in long-term missions such as space gravitational wave detection (e.g., device aging, performance drift, etc.). Future  ground-based experiments will be conducted over extended periods to simulate the complete mission cycle.

To evaluate the charging rate of $V_{\rm TM}$ for long-wavelength UV-LED illumination, experiments were conducted using LED\_265 , LED\_275, LED\_285,  and LED\_295. The UV-LED drive current was set to 10 mA, and the bias electrode was grounded during measurements. Fig. \ref{multi2} shows the $V_{\rm TM}$ variation. By fitting $V_{\rm TM}$ with time using \ref{eq:fit1}, time constants for each wavelength are summarized in Table \ref{tao}. 
As shown in Figure \ref{LED_chara}, the four UV-LEDs exhibit distinct optical powers at a drive current of 10 mA. Given the inverse proportionality between time constants and optical power, Table \ref{tao} also lists the time constants normalized relative to a reference optical power of 1 mW for each UV-LED.

\setcounter{figure}{9}
\setcounter{subfigure}{0}
\begin{subfigure}

\setcounter{figure}{9}
\setcounter{subfigure}{0}
    \begin{minipage}[b]{0.5\textwidth}
        \includegraphics[width=\linewidth]{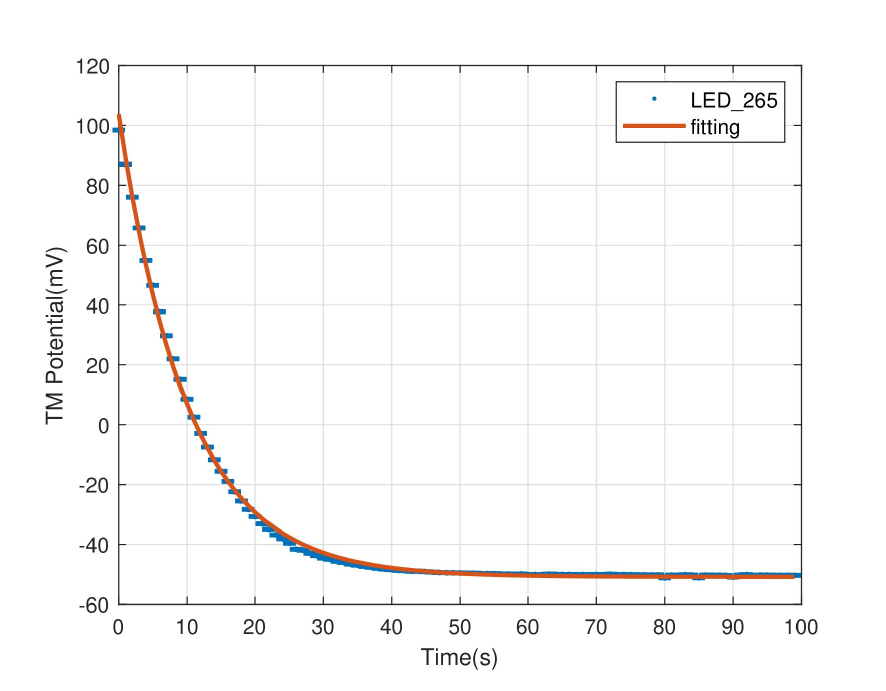}
        \caption{LED\_265.}
        \label{long1}
    \end{minipage}  
\setcounter{figure}{9}
\setcounter{subfigure}{1}
    \begin{minipage}[b]{0.5\textwidth}
        \includegraphics[width=\linewidth]{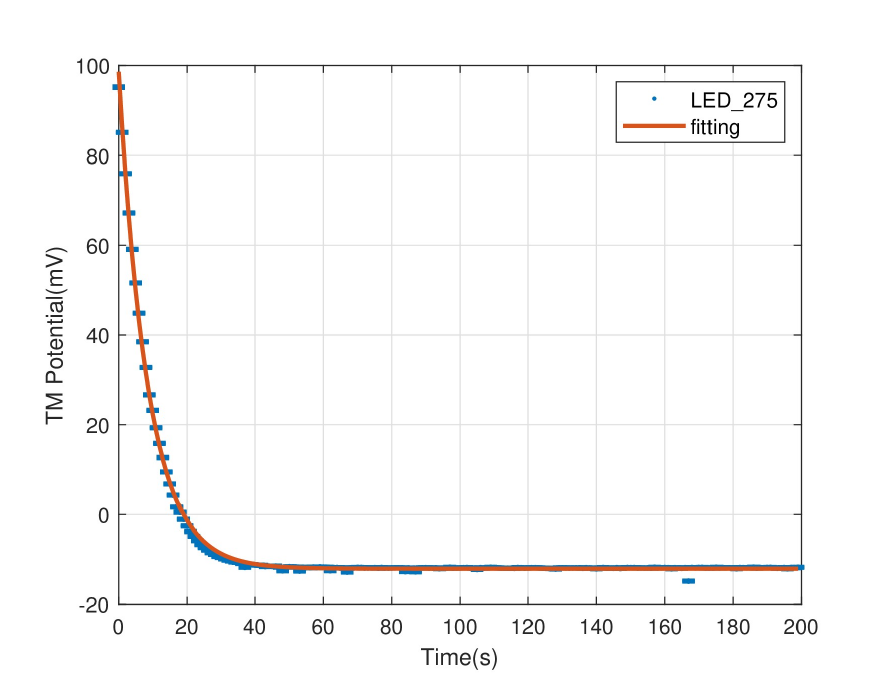}
        \caption{LED\_275.}
        \label{long2}
    \end{minipage}
    
\setcounter{figure}{9}
\setcounter{subfigure}{2}
    \begin{minipage}[b]{0.5\textwidth}
        \includegraphics[width=\linewidth]{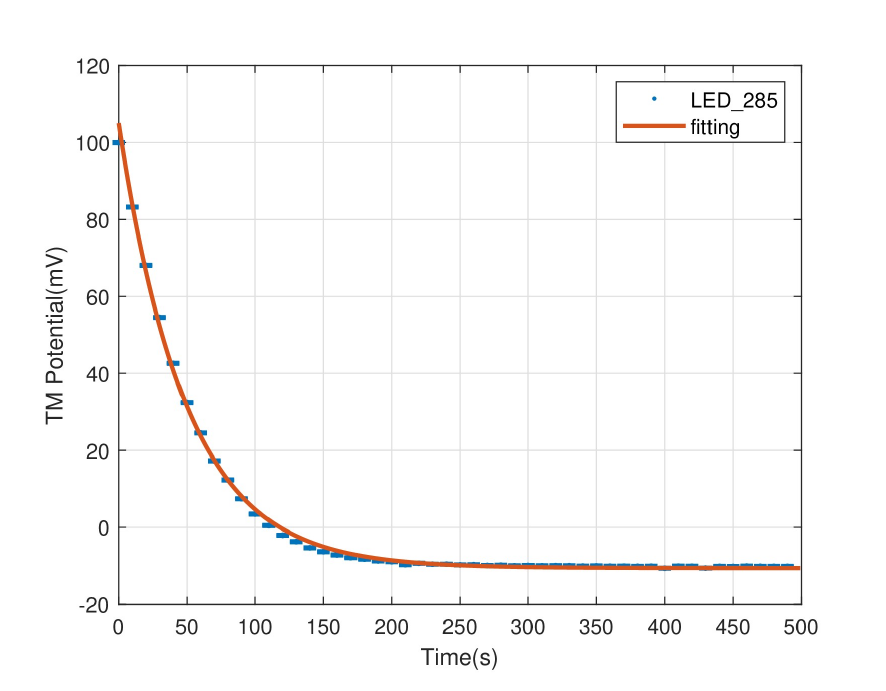}
        \caption{LED\_285.}
        \label{long1}
    \end{minipage}  
\setcounter{figure}{9}
\setcounter{subfigure}{3}
    \begin{minipage}[b]{0.5\textwidth}
        \includegraphics[width=\linewidth]{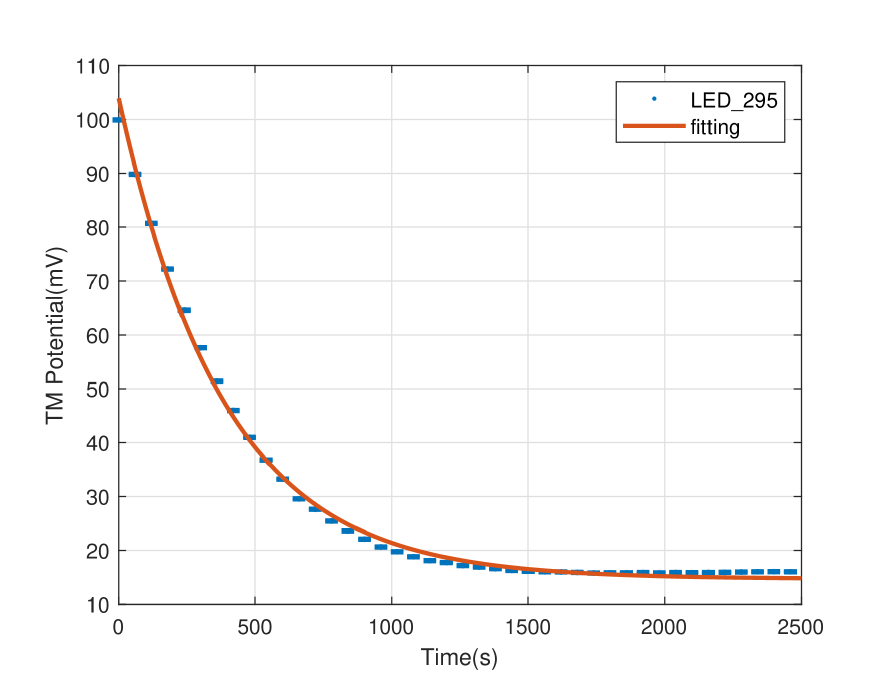}
        \caption{LED\_295.}
        \label{long2}
    \end{minipage}

\setcounter{figure}{9}
\setcounter{subfigure}{-1}
    \caption{TM potential variations due to photoelectrons emitted by UV-LEDs with different wavelengths. \textbf{(A)} LED\_265. \textbf{(B)} LED\_275. \textbf{(A)} LED\_285. \textbf{(B)} LED\_295.}
    \label{multi2}
\end{subfigure}

\begin{table}
\caption{Time constant for different wavelength UV-LEDs.\label{tao}}
\begin{tabular*}{\textwidth}{@{\extracolsep{\fill}}ccc}

\textbf{LED}    &\textbf{Time constant(s)}	& \textbf{Calibrated Time constant(s)}	\\
\hline
LED\_265		& $10.16 \pm 0.08$		& $3.62	\pm 0.08$ \\
LED\_275		& $8.58 \pm 0.04$		& $6.01	\pm 0.07$ \\
LED\_285		& $49.4 \pm 0.2$		& $29.9 \pm 0.4$	\\
LED\_295		& $386 \pm 3$		    & $660 \pm 24$	 \\

\end{tabular*}
\end{table}

As shown in Table \ref{tao}, the time constant after calibration increases with wavelength. The time constants of LED\_285 and LED\_295 are significantly larger than those of LED\_265 and LED\_275. Under conditions normalized to 1 mW output power and identical TM potentials,  LED\_265 and LED\_275 stabilize the potential near equilibrium within 6 seconds. In contrast, LED\_285 and LED\_295 require approximately 30 seconds and over 600 seconds respectively, 5 times and 100 times longer than LED\_265 and LED\_275. This highlights a significant wavelength-dependent efficiency disparity in charge control.

While the preceding experiments utilized DC excitation currents for UV-LED operation, the methodology remains equally applicable in PWM mode. This is critical for space-based gravitational wave detectors, where DC charge control could introduce low-frequency noise into the mHz science band \cite{PMCS}. Fig. \ref{switch}(a) illustrates the potential variation of the TM through three consecutive charge-discharge cycles, employing LED\_275 in PWM mode. The LED was driven at a current of 20 mA and modulated at 1 kHz with 50\% duty cycle. The TM was alternately biased with a positive or negative initial potential, followed by UV-LED illumination until equilibrium was achieved in each case. This alternating sequence was repeated for three cycles, and the shaded area indicates periods of UV-LED activation.

\setcounter{figure}{10}
\setcounter{subfigure}{0}
\begin{subfigure}

\setcounter{figure}{10}
\setcounter{subfigure}{0}
    \begin{minipage}[b]{0.5\textwidth}
        \includegraphics[width=\linewidth]{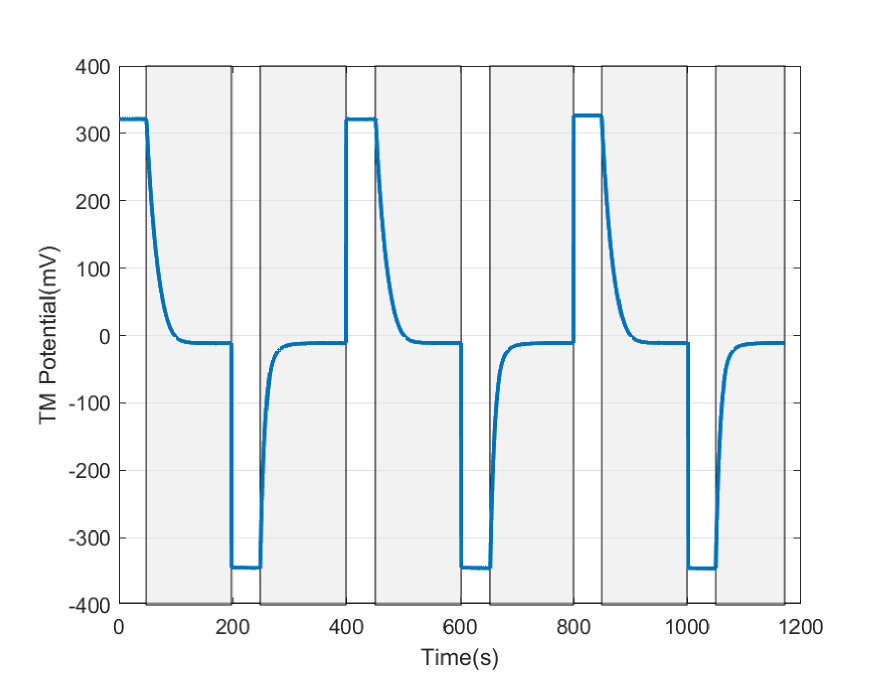}
        \caption{Three charge control cycles.}
        \label{long1}
    \end{minipage}  
\setcounter{figure}{10}
\setcounter{subfigure}{1}
    \begin{minipage}[b]{0.5\textwidth}
        \includegraphics[width=\linewidth]{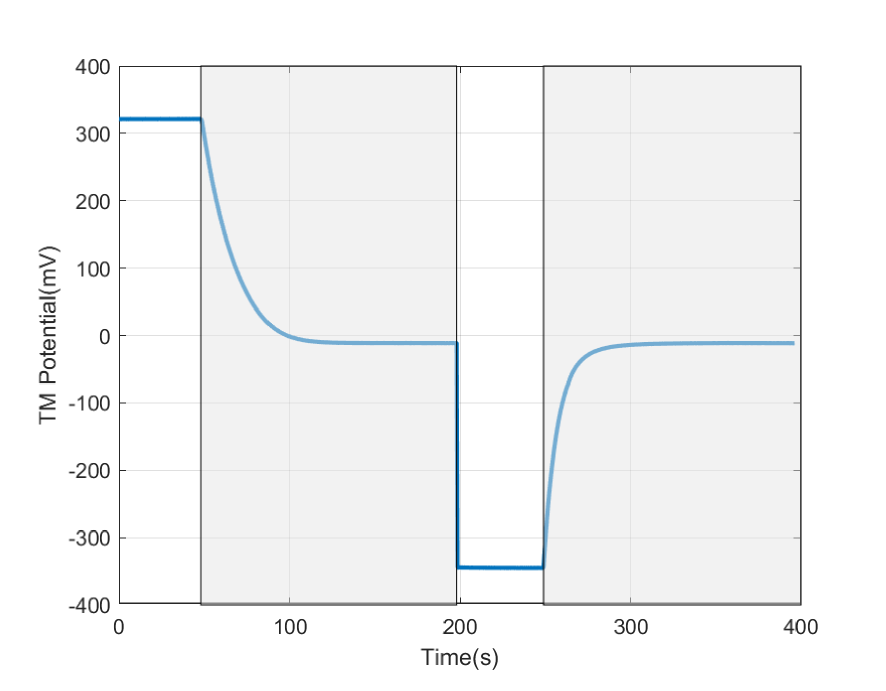}
        \caption{Details of the first cycle.}
        \label{switch}
    \end{minipage}

\setcounter{figure}{10}
\setcounter{subfigure}{-1}
    \caption{Experiments utilized AC excitation currents for UV-LED operation. \textbf{(A)} Three charge control cycles. \textbf{(B)} Details of the first cycle.}
    \label{multi2}
\end{subfigure}

The experimental results demonstrate consistent convergence of the TM potential to equilibrium under UV-LED activation across all three cycles, regardless of initial potential polarity. Fig. \ref{switch}(b) details the transient behavior of the first cycle (0 - 400 s), revealing charge and discharge phases from the initial potentials to equilibrium. This bipolar charge control demonstrates the robustness of the methodology, as demonstrated in the previous research\cite{TATP, ULCC}.

The experimental results demonstrate that for gold-coated surfaces, the LED\_265 showed performance comparable to the LED\_255, and the LED\_275, LED\_285 and LED\_295 exhibit a lower equilibrium potential (within 20 mV) compared to the LED\_255, while wavelengths approaching 300 nm and beyond are ineffective due to insufficient photon energy. Moreover, LED\_275 seems to be an optimal candidate for the source of charge management system, not only achieving a smaller equilibrium potential compared to LED\_255 and LED\_265 but also demonstrating a shorter time constant than long-wavelength UV-LEDs, such as LED\_285 or LED\_295. It should be clarified that our approach does not entirely replace the 255nm UV-LED with the 275nm variant. Instead, it offers more options for neutralizing cosmic ray charging, pending on the solar activity. 
On board, both the 255nm and 275nm should be placed at the four bottom corners of the caging. Depending on the intensity of the solar energetic particles and possibly contamination of the test mass, UV-LED of appropriate wavelengths will be activated for charge management. 

\section{Conclusions.}
Based on a model of inertial sensor that mimics the surface properties and work function of a TM for LISA, UV-LEDs with multiple wavelengths were tested to evaluate their performance in continuous charge management, in terms of both the equilibrium potential and long-term stability. Additionally, the discharge rates (time constant) of UV-LEDs with different wavelengths were measured. The results show that LED\_275 appears to be the best choice if optimizing for both these parameters. These results demonstrate that the 275 nm UV-LED is a promising candidate for the continuous charge management and at the same time for regular rapid discharge operation. 

It is our hope that our work will provide useful input when it comes to the system design and optimisation of the charge management system in the engineering phase of a mission to detect gravitational waves in space.

\section*{Conflict of Interest Statement}

The authors declare that the research was conducted in the absence of any commercial or financial relationships that could be construed as a potential conflict of interest.

\section*{Author Contributions}

Y.K.L., Z.L., and H.H. participated in the design and interpretation of the reported experiments or results. Y.J. and Y. Z. participated in the acquisition and/or analysis of data. Y.J., Y.Z., Z.L., and Y.K.L. participated in drafting and/or revising the manuscript. Y.J., Y.Z., Z.Z., Y.Z., H.L. and S.H. contributed to the experimental data analysis. G.C., H.H., and Z.L. provided administrative, technical, or supervisory support.

\section*{Funding}
This work was  supported by the National Key R\&D Program of China (Task No. 2022YFC2204101).

\section*{Data Availability Statement}
The data used to support the findings of this study are available from the corresponding author upon reasonable request.

\bibliographystyle{Frontiers-Harvard} 
\bibliography{references}


\end{document}